\documentclass[conference,letterpaper]{IEEEtran}

\usepackage[utf8]{inputenc} 
\usepackage[T1]{fontenc}
\usepackage{url}
\usepackage{ifthen}
\usepackage{cite}
\usepackage[cmex10]{amsmath} 

\usepackage{cite}
\ifCLASSINFOpdf
 
\else
   \usepackage[dvips]{graphicx}
 \fi
\usepackage[cmex10]{amsmath}
\usepackage{amssymb}
\usepackage{stfloats}
\usepackage{amsthm} 
\usepackage{wrapfig}
\usepackage{amssymb}
\usepackage{latexsym}
\usepackage{bm, bbm} 
\usepackage{epsfig}
\usepackage{graphicx}
\usepackage{epsfig}
\usepackage{multicol}
\usepackage{psfrag}
\usepackage{float}
\usepackage{cite}
\usepackage{subfigure}
\usepackage{color}

\newcommand{\matr}[1]{{#1}}

\newcommand{\tr}{\mathsf{T}}

\newcommand{\defeq}{\triangleq}

\def\girth{{\rm gir}}

\newtheorem{lemma}{Lemma}
\newtheorem{theorem}[lemma]{Theorem}

\theoremstyle{plain}

\newtheorem{PreDefinition}[lemma]{{\textbf{Definition}}}
    {\begin{PreDefinition}}{\hfill$\square$\end{PreDefinition}}

\theoremstyle{plain}
\newtheorem{Algorithm}[lemma]{Algorithm}

\newtheorem{PreRemark}[lemma]{{\textbf{Remark}}}
  \newenvironment{remark}%
    {\begin{PreRemark}\upshape}{\hfill$\square$\end{PreRemark}}

\newtheorem{PreExample}[lemma]{{\textbf{Example}}}
  \newenvironment{example}%
    {\begin{PreExample}\upshape}{\hfill$\square$\end{PreExample}}

\long\def\symbolfootnote[#1]#2{\begingroup%
\def\thefootnote{\fnsymbol{footnote}}\footnote[#1]{#2}\endgroup} 

\begin{document}
\title{Necessary and Sufficient Girth Conditions for LDPC Tanner Graphs with Denser Protographs} 


                    

 \author{
   \IEEEauthorblockN{Anthony G\'omez-Fonseca\IEEEauthorrefmark{1},
                     Roxana Smarandache\IEEEauthorrefmark{1}\IEEEauthorrefmark{2}, and David G. M. Mitchell\IEEEauthorrefmark{3}}
   \IEEEauthorblockA{
                     Departments of \IEEEauthorrefmark{1}Mathematics and \IEEEauthorrefmark{2}Electrical Engineering,
                     University of Notre Dame,
                     Notre Dame, IN 46556, USA \\
                     \{agomezfo, rsmarand\}@nd.edu}
   \IEEEauthorblockA{\IEEEauthorrefmark{3}%
                     Klipsch School of Electrical and Computer Engineering,
                     New Mexico State University,
                     Las Cruces, NM 88003, USA \\ 
                     dgmm@nmsu.edu}
 }


\maketitle


\begin{abstract}
This paper gives necessary and sufficient conditions for the Tanner graph of a quasi-cyclic (QC) low-density parity-check (LDPC) code based on the all-one protograph to have girth 6, 8, 10, and 12, respectively, in the case of parity-check matrices with column weight 4. These results are a natural extension of the girth results of the already-studied cases of column weight 2 and 3, and it is based on the connection between the girth of a Tanner graph given by a parity-check matrix and the properties of powers of the product between the matrix and its transpose. The girth conditions can be easily incorporated into fast algorithms that construct codes of desired girth between 6 and 12; our own algorithms  are presented for each girth, together with constructions obtained from them and corresponding computer simulations.  More importantly, this paper emphasizes how the girth conditions of the Tanner graph corresponding to a parity-check matrix composed of circulants relate to the matrix obtained by adding (over the integers) the circulant columns of the parity-check matrix. In particular, we show that imposing girth conditions on a parity-check matrix is equivalent to imposing conditions on a square circulant submatrix of size 4 obtained from it. 
\end{abstract}


%

\section{Introduction}\label{sec:Introduction}
Optimized irregular quasi-cyclic (QC) low-density parity-check (LDPC) codes are attractive for implementation purposes due to their algebraic structure that allows for low complexity encoding \cite{lcz+06} and leads to efficiencies in decoder design \cite{wc07}. The performance of an LDPC  code with parity-check matrix $H$ depends on cycles in the associated Tanner graph, since cycles in the
graph cause correlation during iterations of belief propagation decoding. 
Moreover, these cycles form substructures found in the undesirable trapping and absorbing sets that create the  error floor. Cycles have also been shown to decrease the upper bound on the minimum distance (see, e.g., \cite{sv12}). Therefore, codes with large girth are desirable for good performance (large minimum distance and low error floor). Although significant effort has been made to design QC-LDPC code matrices with large minimum distance and girth, e.g., \cite{xcd+07,tss+04,kncs07,phns13,kb13,msc14}, this can be particularly challenging for optimized protographs that contain dense subgraphs, such as those of the AR4JA codes \cite{ddja09} and 5G new radio LDPC codes \cite{richardson2018design}, which contain a significant number of variable nodes with degree larger than 3.

In \cite{sm21}, we have used some previous results by McGowan and Williamson \cite{mw03} and the terminology introduced in  Wu et al. \cite{wyz08}  that  elegantly relate  the girth of $H$ with the girth of $\matr{B}_n(H)\defeq \left(\matr{H}\matr{H}^\tr\right)^{\lfloor{n/2}\rfloor}\matr{H}^{(n\mod 2)}, n\geq 1$,  to highlight the role that certain submatrices of $HH^\tr$ play  in the construction of codes of  desired girth. In particular, we showed that the cycles in the Tanner graph  of a $2N\times n_vN$ parity-check matrix $H$ based on the $(2, n_v)$-regular fully connected (all-one) protograph, with lifting factor $N$, correspond one-to-one to the cycles in the Tanner graph of a $N\times N$ matrix, that we call  $C_{12}$,  obtained from $H$. Similarly, we show that  imposing  girth conditions on a $3N\times n_vN$ parity-check matrix is equivalent to imposing girth conditions on a $3N\times 3N$ submatrix of  $HH^\tr$, which we call $C_H$. 

In order to investigate large girth constructions from dense protographs, this paper extends the results of \cite{sm21} to the case $n_c=4$ and shows how the girth conditions of a $4N\times n_vN$ parity-check matrix are reflected in the corresponding  $4N\times 4N$ submatrix $C_H$ of  $HH^\tr$, and in particular, in a column of $C_H$ given by the sum (over the integers) of the circulant columns of the parity-check matrix.  Although we mostly assume the case  of  an  $(4, n_v)$-regular fully connected protograph,  the results can be used to analyze the girth of the Tanner graph of a  parity-check matrix of zeros and ones. Throughout, we exemplify the techniques and related algorithms by constructing the Tanner graphs of $(4,6)$-regular QC-LDPC codes with girths of 6, 8, 10, and 12, and we conclude the paper with computer simulations of some of the  constructed codes with varying block lengths and girths, confirming the expected robust error control performance.
  
We note that the motivation of the paper is not only to construct good $(4,n_v)$-regular QC-LDPC codes, rather we aim to demonstrate that the approach from \cite{sm21} can be extended to higher column weights and that similar efficient algorithms can be used to construct denser graphs (or sub-graphs) with large girth. As mentioned above, this is particularly important since capacity approaching LDPC codes with irregular protographs often have dense sub-graphs \cite{ddja09}. The necessary and sufficient girth conditions we present here provide a unifying framework for a given girth to be achieved in which all constructions must fit. 
The proposed algorithms to choose lifting exponents are extremely fast, in fact they can be evaluated by hand, and can be used to obtain codes of a given girth for the smallest graph lifting factor $N$. We remark that the technique can be incorporated with other complementary design approaches, such as pre-lifting \cite{msc14} and masking \cite{xcd+07} to construct irregular LDPC codes  that have low error floors from the $(n_c, n_v)$-regular protographs. Finally, note that the technique can also be modified to increase the minimum distance and/or minimum trapping/absorbing set size since cycles appear in the composition of these structures.

\section{Definitions, notations and background}\label{sec:Background}
For any positive integer $L$, let $[L]=\{1,2,\dots,L\}$. An LDPC code $\mathcal{C}$ can be described as the null space of a parity-check matrix $H$, where $\mathcal{C}=\{c\mid Hc^\top=0^\top\}$. We can associate a Tanner graph \cite{tan81} to this matrix $H$ in the usual way.  Its girth, denoted $\girth(H)$, is defined as the length of a shortest cycle in the graph.

A protograph \cite{tho03,ddja09} is a small bipartite graph that can be represented by an $n_c\times n_v$ parity-check or $base$ biadjacency matrix $B=(b_{ij})$, where $b_{ij}>0$ is an integer for each pair $(i,j)$. The parity-check matrix $H$ of a protograph-based LDPC block code can be constructed from $B$ in the following way: each nonzero entry $b_{ij}$ of $B$ is replaced by a summation of $b_{ij}$ non-overlapping permutation matrices of size $N\times N$, and each zero entry is replaced by the $N\times N$ all-zero matrix. In this case we write $H=B^{\uparrow N}$ and $N$ is called the lifting degree. Let $x^a$ (or $I_a$) denote the $N\times N$ circulant permutation matrix obtained from the $N\times N$ identity matrix $I$ by shifting all its entries $a$ positions to the left. 

In this paper we use the triangle operator $\triangle$ introduced in \cite{wyz08} defined as follows. For two nonnegative integers $m$ and $n$, define


{\centering $m\triangle n\triangleq
\begin{cases}
1 & \text{if} \; m\geq2, n=0 \\
0 & \text{otherwise}
\end{cases}.$\\}

\

\noindent This definition can be extended to matrices. Let $M=(m_{ij})$ and $N=(n_{ij})$ be two $s\times t$ matrices. Then we define the $s\times t$ matrix $P=(p_{ij})\triangleq M\triangle N$ entry-wise, where $p_{ij}\triangleq m_{ij}\triangle n_{ij}$ for all pair $(i,j)\in[s]\times[t]$.

The following theorem found in \cite{mw03} and \cite{wyz08} describes an important connection between $\girth(H)$ and matrices $\matr{B}_n(H)\defeq \left(\matr{H}\matr{H}^\tr\right)^{\lfloor{n/2}\rfloor}\matr{H}^{(n\mod 2)}, n\geq 1 $ and offers some insight on the inner structure of the Tanner graph which  simplifies considerably the search for QC protograph-based codes with large girth and minimum distance. 

  
\begin{theorem}(\hspace{-0.01mm}\cite{mw03} and \cite{wyz08})\label{adjacent-cond} A Tanner graph of an LDPC code with parity-check matrix $\matr{H}$ has  $\girth(H)>2g$ if and only if 
 $\matr{B}_t(H)\triangle \matr{B}_{t-2}(H) =\matr{0}, t=2,3,\ldots, g.$ 
  \end{theorem}

\section{Constructing $4\times n_v$ protograph-based QC codes of given girth $g\leq12$}
\label{sec:4xn_v}
In this section, we will construct QC matrices by lifting a $4\times n_v$ protograph, give the conditions required to obtain girth $6\leq g\leq12$, and provide some algorithms to generate these conditions. 

Let $H$ be the parity-check matrix of an $n_cN\times n_vN$, $n_c<n_v$, protograph-based LDPC code given by
\begin{equation}
\label{pcheckmatrix}
H=\left(\begin{matrix} I&I&\cdots&I \\ I&P_{22}&\cdots&P_{2n_v} \\ I&P_{32}&\cdots&P_{3n_v} \\ I&P_{42}&\cdots&P_{4n_v} \end{matrix}\right).
\end{equation}



\noindent For each $i,j\in[4]$, let $P_{i1}=I$ for all $i\in\{2,3,4\}$ and 
\begin{equation}
\label{C_ijmatrices}
C_{ij}=C_{ji}^\top\triangleq P_{i1}P_{j1}^\top+P_{i2}P_{j2}^\top+\cdots+P_{in_v}P_{jn_v}^\top
\end{equation}


\noindent and define
\begin{equation}
C_H\triangleq
\label{polynomialC_H}
\left(\begin{matrix} 0&C_{12}&C_{13}&C_{14} \\ C_{21}&0&C_{23}&C_{24} \\ C_{31}&C_{32}&0&C_{34} \\ C_{41}&C_{42}&C_{43}&0\end{matrix}\right).
\end{equation}
%
%


The following theorem 
characterizes the connection between the matrix $C_H$ and $\girth(H)$,  
derived from the relation established in Theorem \ref{adjacent-cond} between $\girth(H)$ and the matrices $B_n(H)$. 


\begin{theorem}
\label{theorem_girth} 
Let $H$, $C_{ij}$ and $C_H$ be defined as in \eqref{pcheckmatrix}, \eqref{C_ijmatrices} and \eqref{polynomialC_H}, respectively. Then\\
~\\
$\mathrm{\girth}(H)>4 \ \Leftrightarrow  C_H\triangle0=0$, \\
$ {\girth}(H)>6 \  \Leftrightarrow 
C_H\triangle0=0 \ \ \&\   C_HH\triangle H=0 $,\\
$ {\girth}(H)>8\ \Leftrightarrow 
C_H\triangle0=0 \ \ \& \ C_H^2\triangle (I+C_H)=0 $, \\
$\mathrm{\girth}(H)>10 \Leftrightarrow 
\mathrm{\girth}(H)>8 \ \  \&  \ C_H^2H\triangle(H+C_HH)=0$, \\
$\mathrm{\girth}(H)>12 \Leftrightarrow 
{\girth}(H)>10 \ \& \ C_H^3\triangle(I+C_H+C_H^2)=0.$ 
\end{theorem} 
\begin{IEEEproof} Note that  \begin{align*}&\matr{B}_2(H)=\matr{H}\matr{H}^\tr=n_vI+C_H, \quad 
  \matr{B}_3(H)=n_vH+C_HH,  \\
&\matr{B}_4(H)=(n_vI+C_H)^2,   \matr{B}_5(H)=(n_vI+C_H)^2H,  \\&
 \matr{B}_6(H)=(n_vI+C_H)^3, \text{ etc..}
 \end{align*} 
We obtain the following equivalences, completing the proof:\vspace{2mm} $\matr{B}_2(H)\triangle I= 0 \Leftrightarrow C_H\triangle 0=0$;\\ \quad 
 $\matr{B}_3(H)\triangle \matr{B}_1(H)= 0 \Leftrightarrow  C_HH\triangle H=0 $;\\ 
  $B_4(H)\triangle B_2(H) =0 \Leftrightarrow (n_vI+C_H)^2 \triangle  (n_vI+C_H)=0\\\hspace*{0.5cm} \Leftrightarrow C_H^2\triangle (I+C_H)=0;$ \\  $B_5(H)\triangle B_3(H) =0 \Leftrightarrow (n_vI+C_H)^2H \triangle  (n_vI+C_H)H=0\\\hspace*{0.5cm} \Leftrightarrow C_H^2H\triangle (H+C_HH)=0; $  \\  $B_6(H)\triangle B_4(H) =0 \Leftrightarrow (n_vI+C_H)^3 \triangle  (n_vI+C_H)^2=0\\\hspace*{0.5cm} \Leftrightarrow C_H^3\triangle (I+C_H+C_H^2)=0.$\end{IEEEproof}

\begin{remark} Note that, for practical implementation, it is desirable to take each $P_{ij}$ to be a circulant $x^l$, for some $l$, or a permutation matrix lifted to some circulants, for example, $H=({x^l})^{\uparrow N}$. In the remainder of the paper, we consider the first case. The second case was investigated in the case of $n_c=3$ in \cite{msc14,sm21} and is left to future work for $n_c>3$. 
\end{remark}
Suppose that each matrix $P_{ij}$ is a circulant permutation matrix, that is  $P_{2l}=x^{i_l}, P_{3l}=x^{j_l},P_{4l}=x^{k_l},$ for all $l\in [n_v]$, with $i_1=j_1=k_1=0$.   The associated polynomial matrix $H(x)$ is then
\begin{equation}
\label{polynomialpcheckmatrix}
H(x)=\left(\begin{matrix} 1&1&\cdots&1 \\ 1&x^{i_2}&\cdots&x^{i_{n_v}} \\ 1&x^{j_2}&\cdots&x^{j_{n_v}} \\ 1&x^{k_2}&\cdots&x^{k_{n_v}}\end{matrix}\right).
\end{equation} 
\noindent Consider the polynomial matrices $C_{ij}(x)$ and $C_H(x)$ associated with the QC-scalar matrices $C_{ij}$ and $C_H$, then $C_{ij}(x)=C_{ji}^\tr (x)$ for all $i,j\in [n_c]$,
where 
\begin{equation}
\label{polynomialC_ij}
\begin{cases}
\displaystyle C_{21}=\sum_{l=1}^{n_v}x^{i_l},  
\displaystyle C_{31}=\sum_{l=1}^{n_v}x^{j_l},
\displaystyle C_{41}=\sum_{l=1}^{n_v}x^{k_l},\\ 
\displaystyle C_{31}=\sum_{l=1}^{n_v}x^{j_l-i_l},
\displaystyle C_{42}=\sum_{l=1}^{n_v}x^{k_l-i_l}, 
\displaystyle C_{43}=\sum_{l=1}^{n_v}x^{k_l-j_l}.
\end{cases}
\end{equation}


\begin{remark} Note that the transpose of the matrix $$\begin{bmatrix} n_vI & C_{12} &C_{13}&C_{14} \end{bmatrix} $$ is equal to the sum of the $n_v$ circulant columns of $H$ and has an important role in the girth, as we see in Theorem  \ref{theorem_girth}. \end{remark} 
\begin{theorem}
\label{girthH>4}
Let $H(x)$ and $C_H(x)$ be defined as in  \eqref{polynomialpcheckmatrix} and \eqref{polynomialC_H}, respectively. Then ${\girth}(H(x))>4$ if and only if each one of the six sets $\{i_1,i_2,\dots,i_{n_v}\}$, $\{j_1,j_2,\dots,j_{n_v}\}$, $\{k_1,k_2,\dots,k_{n_v}\}$, $\{i_1-j_1,i_2-j_2,\dots,i_{n_v}-j_{n_v}\}$, $\{i_1-k_1,i_2-k_2,\dots,i_{n_v}-k_{n_v}\}$ and $\{j_1-k_1,j_2-k_2,\dots,j_{n_v}-k_{n_v}\}$ contains exactly $n_v$ distinct elements.
\end{theorem}

\begin{IEEEproof} By Theorem \ref{theorem_girth}, $\girth(H)>4$ if and only if $C_H\triangle0=0$. This is equivalent to $C_{ij}(x)\triangle0=0$ for all $1\leq i<j\leq4$. Expanding each of these equations, we obtain
$$ \sum_{l=1}^{n_v}x^{i_l}\triangle0=0, \quad \sum_{l=1}^{n_v}x^{j_l}\triangle 0=0,\quad 
 \sum_{l=1}^{n_v}x^{k_l}\triangle0=0,$$ 
$$ \sum_{l=1}^{n_v}x^{i_l-j_l} \triangle0=0, \quad \sum_{l=1}^{n_v}x^{i_l-k_l} \triangle 0=0, \quad 
\sum_{l=1}^{n_v}x^{j_l-k_l}\triangle0=0.$$

\noindent By using the definition of the triangle operator $\triangle$, we conclude that, for each equation, the exponents should be distinct and the claim follows. \end{IEEEproof}

\

\noindent To choose the exponents $i_l, j_l,$ and $k_l$ satisfying the conditions in Theorem \ref{girthH>4}, we provide the following algorithm to construct a $(4,n_v)$-regular graph with $g>4$. In this algorithm, we first choose $i_1,j_1,k_1$ such that they are not in the specified \emph{forbidden sets}, i.e.,  sets of values that would create  a cycle of size below the desired girth, then we choose $i_2,j_2,k_2$, then we choose $i_3,j_3,k_3$, and so on.

\begin{Algorithm}\label{algorithmgirth4}
(Construct $(4,n_v)$-regular graph with $g>4$)

\noindent Step 1: Set $i_1=0$, $j_1=0$ and $k_1=0$. Set $l=1$.

\noindent Step 2: Let $l:=l+1$. Choose

\hspace{0.2cm} $i_l\notin \{i_m \mid 1\leq m\leq l-1\}$

\hspace{0.2cm} $j_l\notin \{j_m, 
i_l+(j_m-i_m) \mid 1\leq m\leq l-1\}$

\hspace{0.2cm} $k_l\notin \{k_m,  
i_l+(k_m-i_m), 
j_l+(k_m-j_m) \mid 1\leq m\leq l-1\}$

\noindent Step 3: If $l=n_v$ stop; otherwise go to Step 2.
\end{Algorithm}


\begin{example}
\label{example1_girth4}
In this example, we construct a $4\times6$ protograph-based matrix using Algorithm \ref{algorithmgirth4}. In each iteration $l$, $2\leq l\leq n_v$, we choose the smallest positive value for each of $i_l,j_l,$ and $k_l$. We obtain 


{\centering $H(x)=\left(\begin{matrix} 1&1&1&1&1&1 \\ 1&x&x^2&x^3&x^4&x^5 \\ 1&x^2&x&x^5&x^7&x^3 \\ 1&x^3&x^5&x&x^9&x^2 \end{matrix}\right).$ \\}


\noindent If we choose lifting degree $N=\displaystyle \left(\max_{1\leq l\leq6}\{i_l,j_l,k_l\}\right)+1=10$, then $H(x)$ has girth 6. 
\end{example}


\begin{example}
\label{example2_girth4}
Using Algorithm \ref{algorithmgirth4}, we construct a protograph-based matrix as in Example \ref{example1_girth4}. In each iteration $l$, $2\leq l\leq n_v$, however, we choose each of $i_l,j_l,$ and $k_l$ to be one more than the maximum value in their corresponding forbidden sets. We obtain 
\begin{equation}\label{fan} 
H(x)=\left(\begin{matrix} 1&1&1&1&1&1 \\ 
1&x&x^2&x^3&x^4&x^5 \\ 
1&x^2&x^4&x^6&x^8&x^{10} \\ 
1&x^3&x^6&x^9&x^{12}&x^{15} 
\end{matrix}\right).
\end{equation}

\noindent If we let $N=\displaystyle \left(\max_{1\leq l\leq6}\{i_l,j_l,k_l\}\right)+1=16$, then the girth of $H(x)$ is 6. Notice that $N=16$ is not the smallest positive value for which $\girth(H(x))>4$. If we choose $N=7$, then $\girth(H(x))=6$. We note that \eqref{fan} is a shortened version of Fan's array construction \cite{fan00} that gives $g=6$ for $N=7$.
\end{example}


\begin{theorem}
\label{girthH>6}
Let $H(x)$ and $C_H(x)$ be defined as in equations \eqref{polynomialpcheckmatrix} and \eqref{polynomialC_H}, respectively. Then $\girth(H(x))>6$ if and only if, for any $m\in[n_v]$, each one of these four sets 
contains distinct elements:
\


$\{i_m-i_n, j_m-j_n, k_m-k_n  \mid n\in[n_v], n\neq m\}$, 

\

$\{i_n,  
i_n-j_n+j_m,  
i_n-k_n+k_m,  
i_p+(j_n-j_m)+(k_m-k_p) \mid \\ 
~ ~ ~ ~  n,p\in[n_v], p\neq m,n\neq m\}$,

\

$\{j_n, 
j_n-i_n+i_m,  
j_n-k_n+k_m, 
j_p+(i_n-i_m)+(k_m-k_p) \mid\\~ ~ ~ ~  n,p\in[n_v], p\neq m,n\neq m\}$,

\

$\{k_n,  
k_n-i_n+i_m, 
k_n-j_n+j_m, 
k_p+(i_n-i_m)+(j_m-j_p) \mid \\~ ~ ~ ~ n,p\in[n_v], p\neq m,n\neq m\}$.
\end{theorem}

\begin{Algorithm}
\label{algorithmgirth6}
(Construct $(4,n_v)$-regular graph with $g>6$)
\

%
%
\noindent Step 1: Set $i_1=0$, $j_1=0$ and $k_1=0$. Set $l=1$.

\noindent Step 2: Let $l:=l+1$. Choose

\noindent $i_l\notin \{i_m, (i_m-j_m)+j_n, (j_m-k_m)+(k_n-j_n)+i_p, (i_m-k_m)+k_n, (k_m-j_m)+(j_n-k_n)+i_p \mid 1\leq m,n,p\leq l-1\}$

\noindent $j_l\notin \{j_m, i_l+j_m-i_n, i_m+(j_n-i_n), (i_m-k_m)+(k_n-i_n)+j_p, i_l+(k_m-i_m)+(j_n-k_n), (j_m-k_m)+k_n, i_l+(j_m-k_m)+k_n-i_p, 2i_l+(k_m-i_m)+(j_n-k_n)-i_p \mid 1\leq m,n,p\leq l-1\}$

\noindent $k_l\notin \{k_m, j_l+k_m-j_n, i_l+k_m-i_n, j_l+(i_m-j_m)+(i_n-k_n), i_m+(k_n-i_n), j_l+i_m-j_n+(k_p-i_p), 2j_l+(i_m-j_m)+(k_n-i_n)-j_p, (k_m-j_m)+j_n, i_l+(j_m-i_m)+(k_n-j_n), i_l+(k_m-j_m)+j_n-i_p, 2i_l+(j_m-i_m)+(k_n-j_n)-i_p \mid 1\leq m,n,p\leq l-1\}$

\noindent Step 3: If $l=n_v$ stop; otherwise go to Step 2.
\end{Algorithm}

\begin{example}
\label{example1_girth6}
In this example, we construct a $4\times6$ protograph-based matrix using Algorithm \ref{algorithmgirth6}. In each iteration $l$, $2\leq l\leq n_v$, we choose the smallest positive value for each of $i_l,j_l,$ and $k_l$ as we did in Example \ref{example1_girth4}. We obtain
\begin{equation}\label{g8N85}
    H(x)=\left(\begin{matrix} 1&1&1&1&1&1 \\ 1&x&x^5&x^8&x^{10}&x^{25} \\ 1&x^3&x^{14}&x^{29}&x^{49}&x^{96} \\ 1&x^4&x^2&x^{36}&x^{55}&x^{108} \end{matrix}\right).
\end{equation}

\noindent If we choose lifting degree $N=\displaystyle \left(\max_{1\leq l\leq6}\{i_l,j_l,k_l\}\right)+1=109$, then $H(x)$ has girth 8. The smallest positive integer $N$ required to obtain $\girth(H(x))>6$ is $N=85$. Simulation results are provided for  codes obtained from \eqref{g8N85} with $N=85$ and $N=347$ in Section \ref{sec:sim}.
\end{example}


\begin{example}\label{example2_girth6}
If we choose values of $i_l,j_l,$ and $k_l$ one more than the maximum value in their corresponding forbidden sets (instead of choosing the smallest positive value for each of $i_l,j_l,$ and $k_l$, as in Example \ref{example1_girth6}) we obtain the following matrix
\begin{equation*}
H(x)=\left(\begin{matrix} 1&1&1&1&1&1 \\ 1&x&x^8&x^{54}&x^{355}&x^{2324} \\ 1&x^3&x^{23}&x^{154}&x^{1011}&x^{6617} \\ 1&x^7&x^{53}&x^{354}&x^{2323}&x^{15203} \end{matrix}\right).
\end{equation*}

\noindent For these circulants, $N=111$ is the smallest value that can obtain this girth. 

Reducing the exponents modulo 111, we obtain 
\begin{equation}\label{g8N111}
    H(x)=\left(\begin{matrix} 1&1&1&1&1&1 \\ 1&x&x^8&x^{54}&x^{22}&x^{104} \\ 1&x^3&x^{23}&x^{43}&x^{12}&x^{68} \\ 1&x^7&x^{53}&x^{21}&x^{103}&x^{107} \end{matrix}\right),
\end{equation}
which also has girth 8 for $N=111$. Note that the smallest positive integer to obtain girth 8 in \eqref{g8N111} is now $N=105$. Simulation results are provided for  codes obtained from \eqref{g8N111} with $N=111$ and $N=347$ in Section \ref{sec:sim}.
\end{example}


\begin{theorem}
\label{girthH>8}
Let $H(x)$ and $C_H(x)$ be defined as in  \eqref{polynomialpcheckmatrix} and \eqref{polynomialC_H}, respectively. Then $\girth(H(x))>8$ if and only if $\girth(H(x))>4$ and each one of these sixteen sets contains distinct elements, for all $u,w\in [n_v], u\neq w$:
 \\ ~\\ $\{(i_u-j_u)+j_w,(i_u-k_u)+k_w 
\}, \{(j_u-i_u)+i_w,(j_u-k_u)+k_w 
\}$,
\\
 $\{(k_u-i_u)+i_w,(k_u-j_u)+j_w 
\},  \{(j_u-i_u)-j_w,(k_u-i_u)-k_w 
\}$, 
\\
$\{j_u-i_w,(j_u-k_u)-(i_w-k_w) 
\}, \{k_u-i_w,(k_u-j_u)-(i_w-j_w) 
\}$, 
\\
$\{(i_u-j_u)-i_w,(k_u-j_u)-k_w 
\}, \{i_u-j_w,(i_u-k_u)-(j_w-k_w) 
\},$ 
\\
$\{k_u-j_w,(k_u-i_u)-(j_w-i_w) 
\}, \{(i_u-k_u)-i_w,(j_u-k_u)-j_w 
\},$ 
 \\
$\{i_u-k_w,(i_u-j_u)-(k_w-j_w) 
\},\{i_u-i_w,j_u-j_w,k_u-k_w 
\}, $ 
 \\  $\{j_u-j_w,(i_u-j_u)-(i_w-j_w),(j_u-k_u)-(j_w-k_w)\}$, 
\ \\ 
$\{k_u-k_w,(i_u-k_u)-(i_w-k_w),(j_u-k_u)-(j_w-k_w)
\}$,
\\ $\{i_u-i_w,(i_u-j_u)-(i_w-j_w),(i_u-k_u)-(i_w-k_w)\},$
$\{j_u-k_w,(j_u-i_u)-(k_w-i_w) 
\}.$

\end{theorem}


\begin{example}
\label{example1_girth8}
We construct a $4\times6$ matrix $H$ using an algorithm derived from Theorem \ref{girthH>8}, where, at each iteration $l$, $l\in [n_v],$  we choose the smallest possible positive value for each of $i_l,j_l, k_l$, as we did in Examples \ref{example1_girth4} and \ref{example1_girth6}.\footnote{Due to space constraints, we omit algorithms corresponding to Theorems 13 and 15; they can be written in the same way as Algorithms 6 and 10.} We obtain
 $$H(x)=\left(\begin{matrix} 1&1&1&1&1&1 \\ 1&x&x^9&x^{28}&x^{41}&x^{75} \\ 1&x^3&x^{21}&x^{54}&x^{98}&x^{180} \\ 1&x^7&x^{38}&x^{93}&x^{162}&x^{297} \end{matrix}\right).$$
If we choose $N=\displaystyle 2\left(\max_{1\leq l\leq6}\{i_l,j_l,k_l\}\right)+1=595$, then $H(x)$ has girth 10. The smallest $N$ required to obtain $\girth(H(x))>8$ is $N=347$. The resulting code is simulated in Section \ref{sec:sim}.
\end{example}


\begin{theorem}
\label{girthH>10}
Let $H(x)$ and $C_H(x)$ be defined as in equations \eqref{polynomialpcheckmatrix} and \eqref{polynomialC_H}, respectively. Then $\girth(H(x))>10$ if and only if $\girth(H(x))>8$ and, for any $l\in[n_v]$, each one of these four sets contains distinct elements: \\~\\
 $\{i_u-i_w, j_u-j_w, k_u-k_w, i_l+(j_u-i_u)-j_w, i_l+(k_u-i_u)-k_w, j_l+(i_u-j_u)-i_w, j_l+(k_u-j_u)-k_w, k_l+(i_u-k_u)-i_w, k_l+(j_u-k_u)-j_w \mid u,w\in[n_v], u\neq w, u\neq l\}$,\\~\\
 $\{(i_u-j_u)+j_w, (i_u-k_u)+k_w, i_l+i_u-i_w, i_l+(i_u-j_u)-(i_w-j_w), i_l+(i_u-k_u)-(i_w-k_w), j_l+i_u-j_w, j_l+(i_u-k_u)-(j_w-k_w), k_l+i_u-k_w, k_l+(i_u-j_u)-(k_w-j_w) \mid u,w\in[n_v], u\neq w, w\neq l\}$,\\~\\
$\{(j_u-i_u)+i_w, (j_u-k_u)+k_w, i_l+j_u-i_w, i_l+(j_u-k_u)-(i_w-k_w), j_l+j_u-j_w, j_l+(i_w-j_w)-(i_u-j_u), j_l+(j_u-k_u)-(j_w-k_w), k_l+j_u-k_w, k_l+(j_u-i_u)-(k_w-i_w) \mid u,w\in[n_v], u\neq w, w\neq l\}$,\\~\\
 $\{(k_u-i_u)+i_w, (k_u-j_u)+j_w, i_l+k_u-i_w, i_l+(k_u-j_u)-(i_w-j_w), j_l+k_u-j_w, j_l+(k_u-i_u)-(j_w-i_w), k_l+k_u-k_w, k_l+(i_w-k_w)-(i_u-k_u), k_l+(j_w-k_w)-(j_u-k_u) \mid u,w\in[n_v], u\neq w, w\neq l\}.$ \end{theorem}
\begin{example}
\label{example1_girth10}
In this example, we construct a $4\times6$ protograph-based matrix using an algorithm derived from Theorem \ref{girthH>10}. In each iteration $l$, $2\leq l\leq n_v$, we choose the smallest positive value for each of $i_l,j_l$ and $k_l$. We obtain
$$H(x)=\left(\begin{matrix} 1&1&1&1&1&1 \\ 1&x&x^{12}&x^{45}&x^{147}&x^{445} \\ 1&x^3&x^{31}&x^{126}&x^{320}&x^{980} \\ 1&x^7&x^{67}&x^{231}&x^{636}&x^{1626} \end{matrix}\right).$$
If we choose $N=\displaystyle 2\left(\max_{1\leq l\leq6}\{i_l,j_l,k_l\}\right)+1=3253$, then $H(x)$ has girth 12, however, the smallest $N$ required to obtain $\girth(H(x))>10$ is $N=1881$.  The resulting code is simulated in Section \ref{sec:sim}.
\end{example}
\begin{remark} We note that we could also use a computer to search for the possible values in the same way, one by one, with techniques such as progressive edge growth (PEG) \cite{hea05}, but the last values in the matrix are hard to obtain, particularly as the density of the protograph increases. However, the proposed algorithms immediately give the next possible value and can be modified to return the size $N$ needed in a similar way to the formulation in \cite{sm21}. The algorithms can also be modified so that a random value among the possible is chosen at each time in order to optimize the performance. Or it can be chosen such that the smallest possible value can be taken at each point so that the smallest $N$ is obtained. If a choice is not possible at some point for a desired $N$, backtracking can be added to pick a different value at a previous step, until a value is available at the current step.  Finally, we note that the algorithm can also be modified to increase the minimum distance and/or minimum trapping/absorbing set size since cycles appear in the composition of these structures. \end{remark}

\section{Simulation results}\label{sec:sim}
To verify the performance of the constructed codes, computer simulations were performed assuming binary phase shift keyed (BPSK) modulation and an additive white Gaussian noise (AWGN) channel. The sum-product message passing decoder was allowed a maximum of 100 iterations and employed a syndrome-check based stopping rule. 

\begin{figure}[t]
\begin{center}
\includegraphics[width=\columnwidth]{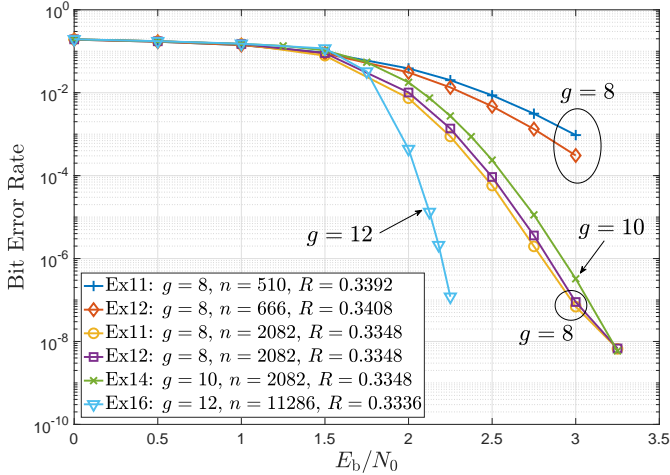}\vspace{-3mm}
\end{center}
\caption{Simulated decoding performance in terms of BER for the $R\approx 1/3$, $(4,6)$-regular QC-LDPC codes from Examples 11-16.}\label{fig:anthonysim}\vspace{-4mm}
\end{figure}

In Fig.~\ref{fig:anthonysim}, we plot the bit error rate (BER) for several $R\approx 1/3$ $(4,6)$-regular QC-LDPC codes from Examples 11-16 with varying code lengths and girth.  For comparison, we selected a larger lifting factor than the minimum for the codes from Examples 11 and 12 with  ($N=347$ corresponding to block length $n=2082$, both codes retain $g=8$) to match the block length of the $g=10$ code from Example 14. We note that the girth $g=8$ codes have similar performance, and a slightly better waterfall, than the girth $10$ code, but they also display the beginning of an error floor at $3.25$dB. The Example 14 and 16 codes with girth $g=10$ and $g=12$, respectively, display no indication of an error-floor, at least down to respective BERs of $10^{-8}$ and $10^{-7}$. The Example 16 code with $g=12$ has a larger lifting factor  $N=1881$ and the resulting code with block length $n=11286$ shows a waterfall approximately $0.58$dB from the iterative decoding threshold ($1.67$dB) for $(4,6)$-regular LDPC codes  \cite{ru08} at a BER of $10^{-7}$.

\section{Concluding remarks}\label{sec:conc}
In this paper we gave necessary and sufficient conditions for the Tanner graph of a protograph-based QC-LDPC code with column weight 4 to have girth $6\leq g\leq 12$, successfully extending the approach of \cite{sm21} to denser protographs. The girth conditions were used to write fast algorithms which were exemplified by constructing the Tanner graphs of $(4,6)$-regular QC-LDPC codes with girths of 6, 8, 10, and 12. The necessary and sufficient girth conditions we presented provide a unifying framework for a given girth to be achieved in which all constructions must fit. Obtaining large girth for relatively dense graphs is a challenging and important topic since capacity approaching irregular LDPC codes often have such sub-graphs in the protograph. Future work involves extending the techniques in this paper to optimized irregular protographs to achieve large girth and low error floors.

\section*{Acknowledgements}
This material is based upon work supported by the National Science Foundation under Grant Nos. OIA-1757207 and HRD-1914635. A. G. F. thanks the support of the GFSD (formerly NPSC) and Kinesis-Fern\'andez Richards fellowships.

\bibliographystyle{IEEEtran}

\end{document}